\documentclass[10pt,letterpaper]{article}
\usepackage{opex3}
\usepackage{color}
\usepackage{epsf}
\usepackage{cite}
\usepackage{bm}

\begin{document}

\title{Plasmon enhanced Raman scattering effect for an atom near a carbon nanotube}

\author{I.V.~Bondarev\large$^\ast$}

\address{Math \& Physics Department, North Carolina Central University, 1801 Fayetteville Str, Durham, NC 27707, USA}

\hskip4.87cm\small{$\ast$}\\[-0.57cm]\email{ibondarev@nccu.edu}

\begin{abstract}
Quantum electrodynamics theory of the resonance Raman scattering is developed for an atom in a close proximity to a carbon nano\-tube. The theory predicts a dramatic enhancement of the Raman intensity in the strong atomic coupling regime to nanotube plasmon near-fields. This resonance scattering is a manifestation of the general electromagnetic surface enhanced Raman scattering effect, and can be used in designing efficient nanotube based optical sensing substrates for single atom detection, precision spontaneous emission control, and manipulation.
\end{abstract}

\section{Introduction}
Surface-enhanced Raman spectroscopy (SERS) has received much of attention recently due to a very broad range of its applications ranging from optics and plasmonics to biochemistry and medicine~\cite{Moscovits01,Otto05,ZhangNat13}. High scattering intensities within narrow spectral bands reduce the probability for spectral overlapping to allow for better recognition of multiple markers, making SERS one of the most efficient optical sensing techniques. With the development of advanced nanomaterials, various SERS substrates are demonstrated~\cite{Peng,Pimenta,Hao,Lv,Lin,Chen,Sun}. However, there is still a need for inexpensive substrates of improved sensitivity and signal reproducibility, which require clear understanding of the underlying scattering mechanisms to be developed.

In general, the SERS effect originates from the resonance increase of the induced transition dipole moment of an atomic or molecular scatterer when positioned in the near-surface zone of a metallic structure. This can be for two reasons. They are: (i)~due to quasi-static electric fields associated with resonance plasmon excitation in metallic structures, and/or (ii)~due to the electron polarizability increase associated with the charge transfer between the substrate and the scatterer. It is generally agreed to distinguish between the electromagnetic (EM) and chemical SERS effects, accordingly~\cite{Moscovits01,Otto05,ZhangNat13}.

Most of the applications of carbon nanotubes (CNs) to increase the Raman scattering signal have been to decorate them with metallic nanoparticles~\cite{Chen,Sun}, in order to obtain local EM field enhancement generated by the spatially confined plasmon modes of the nanoparticles with CNs only used as a network to support the particles. Only very recently, the chemical SERS effect with CNs alone, without any combination with metallic nanoparticles, was first reported by Andrada et al. in~\cite{Pimenta}, where molecules covalently bound to single wall CNs demonstrated the resonantly increased Raman signal that was even stronger than that of the nanotube itself.

\begin{figure}
\epsfxsize=8.0cm\centering{\epsfbox{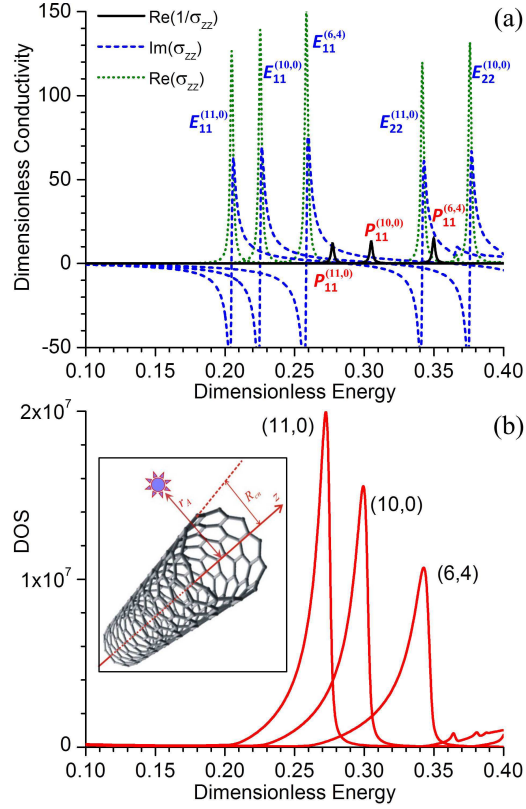}}\caption{(a)~Fragment of the energy dependence of the dimensionless (normalized by $e^2/2\pi\hbar$) axial surface conductivities $\sigma_{zz}$ for the semiconducting (6,4), (10,0) and (11,0) nanotubes of increasing diameter. Peaks of $\mbox{Re}\,\sigma_{zz}$ represent excitons ($E_{11}$, $E_{22}$, ...); peaks of $\mbox{Re}(1/\sigma_{zz})$ are inter-band plasmons ($P_{11}$, ...). (b)~Photonic DOS functions for the CNs in (a) with the TLS placed at the distance $r_A\!=\!R_{cn}+2b$ (see inset). Dimensionless energy is $[Energy]/2\gamma_0$. Conductivities are obtained using the ($\mathbf{k}\cdot\mathbf{p}$)-scheme developed by Ando~\cite{Ando}. DOS functions are calculated as described by Bondarev and Lambin in~\cite{Bondarev05,Bondarev06}. See text for notations.}\label{fig1}
\end{figure}

In this article, a quantum electrodynamics (QED) theory of the resonance Raman scattering is presented for an atom near a carbon nanotube, to demonstrate that individual CNs are capable of providing the electromagnetic SERS effect as well. Nanotubes offer extraordinary stability, flexibility and precise tunability of their EM properties on-demand by simply varying their diameters and/or chiralities. CNs of different diameters and chiralities feature similar electronic band structure peculiarities, yet shifted in frequency relative to one another~\cite{Saito,Ando,ChemPhysSI}. This yields similar EM properties over a broad range of excitation frequencies both in the far- and in the near-field zone, originating from exciton and plasmon excitations, respectively. Excitons and plasmons are different in their physical nature, but originate from the same circumferentially quantized electronic transitions. Due to the circumferential quantization of the longitudinal electron motion, real axial (along the CN axis) optical conductivities of single wall CNs consist of series of peaks $E_{11}, E_{22}, ...$, representing the 1st, 2nd, etc. excitons, respectively [see Fig.~\ref{fig1}(a)]. Imaginary conductivities are linked with the real ones by the Kramers-Kronig relation, and so real \emph{inverse} conductivities show the resonances $P_{11}, P_{22}, ...$ [Fig.~\ref{fig1}(a)] next to their excitonic counterparts.~These are inter-band plasmons that were theoretically demonstrated quite recently to play the key role in a variety of new interesting surface EM phenomena with CNs~\cite{Bondarev09,Popescu11,BondarevNova11,Bondarev12,Bondarev12pss,Bondarev06PRB,Bondarev07,GelinBondarev13,Woods13,Bondarev14,GelinBondarev14}, including exciton-plasmon coupling~\cite{Bondarev09,BondarevNova11} and plasmon generation by excitons~\cite{Bondarev12,Bondarev12pss}, exciton Bose-Einstein condensation in individual single wall CNs~\cite{Bondarev14}, Casimir attraction in double wall CNs~\cite{Popescu11,BondarevNova11,Woods13}, resonance optical absorption~\cite{Bondarev06PRB} and atomic entanglement in hybrid systems of extrinsic atoms/ions doped into CNs~\cite{Bondarev07,GelinBondarev13,GelinBondarev14}, to mention a few, --- all of direct relevance to conceptually new tunable optoelectronic device applications with carbon nanotubes~\cite{Bondarev10JCTN,Bondarev14Dekker}. Experimental evidence for these low-energy ($\sim\!1\!-\!2$~eV) weakly-dispersive plasmon modes in CNs was first reported by Pichler et al. in~\cite{Pichler98}.~Inter-band plasmons are \emph{standing} charge density waves due to the periodic opposite-phase displacements of the electron shells with respect to the ion cores in the neighboring elementary cells of the CN~\cite{Bondarev12,Bondarev12pss}. When excited, their (plasmon-induced) quasi-static electric fields can be strong enough to result in the enhanced Raman scattering effect by atomic type species (extrinsic atoms, ions, molecules, or semiconductor quantum dots) in the CN vicinity.~This work derives and analyzes the differential cross-section for such scattering.

\section{The Hamiltonian}

In absence of external EM radiation, an atom, modeled here by a two-level system (TLS) positioned at the point $\mathbf{r}_{A}$ near an infinitely long single wall CN, interacts with the \emph{quantum} EM field of the CN via an electric transition dipole moment $d_{z}\!=\!\langle u|\hat{d}_z|l\rangle$ between the TLS lower and upper states, $|l\rangle$ and $|u\rangle$, respectively, with the $z$-quantization axis being the CN symmetry axis [Fig.~\ref{fig1}(b), inset]. Transverse dipole orientations can be neglected due to the strong transverse depolarization effect in individual CNs~\cite{Tasaki98,Li01,Bondarev02,Marin03,Bondarev04}. The full QED second quantized Hamiltonian for such a coupled CN--TLS quantum system was earlier formulated by Bondarev and Lambin in~\cite{Bondarev05,Bondarev06} to have the following form
\begin{eqnarray}
\hat{H}\!\!\!&=&\!\!\!\hat{H}_F+\hat{H}_A+\hat{H}_{AF}=\int_{0}^{\infty}\!\!\!\!\!d\omega\,\hbar\omega\!\int\!d\mathbf{R}\,\hat{f}^{\dag}(\mathbf{R},\omega)\hat{f}(\mathbf{R},\omega)+
{\hbar\tilde{\omega}_{A}\over{2}}\,\hat{\sigma}_{z}\nonumber\\
\!\!\!&+&\!\!\!\int_{0}^{\infty}\!\!\!\!\!d\omega\!\int\!d\mathbf{R}\;[\,\mbox{g}^{(+)}(\mathbf{r}_{A},\mathbf{R},\omega)\,\hat{\sigma}^{\dag}
-\mbox{g}^{(-)}(\mathbf{r}_{A},\mathbf{R},\omega)\,\hat{\sigma}\,]\hat{f}(\mathbf{R},\omega)+\mbox{h.c.}\,,\label{Htwolev}
\end{eqnarray}
with the three terms representing the (medium-assisted) quantum EM field of the CN, the TLS, and their interaction, respectively. Here, $\hat{f}^{\dag}(\mathbf{R},\omega)$ and $\hat{f}(\mathbf{R},\omega)$ are the scalar bosonic field operators that create and annihilate, respectively, surface EM excitations of frequency $\omega$ in the CN field subsystem, $\,\mathbf{R}\!=\!(R_{cn},\varphi,z)$ is the radius-vector of a point on the CN surface. Pauli operators, $\hat{\sigma}_{z}\!=\!|u\rangle\langle u|-|l\rangle\langle l|$, $\hat{\sigma}\!=\!|l\rangle\langle u|$ and $\hat{\sigma}^{\dag}\!=\!|u\rangle\langle l|$, describe the TLS and its electric dipole transitions between the two states, upper $|u\rangle$ and lower $|l\rangle$, with the transition frequency $\omega_{A}$ modified by the diamagnetic $\hat{\mathbf{A}}^{2}$-term (vector potential) to result in the new \emph{renormalized} transition frequency
$\tilde{\omega}_{A}=\omega_{A}[1-2/(\hbar\omega_{A})^{2}\!\int_{0}^{\infty}\!d\omega\!\int\!d\mathbf{R}|\mbox{g}^{\perp}(\mathbf{r}_{A},\mathbf{R},\omega)|^{2}]$. The matrix elements of the CN field interaction with the TLS are of the form $\mbox{g}^{(\pm)}\!=\!\mbox{g}^{\perp}\pm(\omega/\omega_{A})\mbox{g}^{\parallel}$~with $\mbox{g}^{\perp(\parallel)}(\mathbf{r}_{A},\mathbf{R},\omega)\!=
\!-i(4\omega_{A}/c^{2})\sqrt{\pi\hbar\omega\,\mbox{Re}\,\sigma_{zz}(\omega)}\,\,d_{z}^{\;\,\perp(\parallel)}G_{zz}(\mathbf{r}_{A},\mathbf{R},\omega)$,~where $^{\perp(\parallel)}G_{zz}$ is the $zz$-component of the transverse (longi\-tudinal) Green tensor (with respect to the first variable) of the CN assisted quantum field, $\sigma_{zz}(\omega)$ is the CN surface axial conductivity [Fig.~\ref{fig1}(a)]. Functions $\mbox{g}^{(\pm)}$ have the property as follows,
\begin{equation}
\int\!d\mathbf{R}\,|\mbox{g}^{(\pm)}(\mathbf{r}_{A},\mathbf{R},\omega)|^{2}\!=\!\frac{\hbar^2}{2\pi}\Gamma_{0}(\omega)\!
\left[\xi^\parallel(\mathbf{r}_{A},\omega)+\frac{\omega_{A}^2}{\omega^{2}}\xi^\perp(\mathbf{r}_{A},\omega)\right],
\label{gpm}
\end{equation}
where $\xi^{\perp(\parallel)}(\mathbf{r}_{A},\omega)\!=\!\mbox{Im}^{\perp(\parallel)}G_{zz}^{\,\perp(\parallel)}(\mathbf{r}_{A},\mathbf{r}_{A},\omega)/\mbox{Im}G_{zz}^{0}(\omega)$
is the transverse (longitudinal) photonic density of states (DOS) relative to vacuum as seen from the TLS location $\textbf{r}_A$, and $\Gamma_{0}(\omega)\!=\!8\pi\omega^{2}d_{z}^{2}\,\mbox{Im}G_{zz}^{0}(\omega)/\hbar c^{2}$ is the TLS spontaneous decay rate in vacuum with $\mbox{Im}\,G_{zz}^{0}(\omega)\!=\!\omega/6\pi c$ being the vacuum imaginary Green tensor $zz$-component.

Hamiltonian~(\ref{Htwolev}) involves only two standard approximations, the electric dipole and two-level approximation~\cite{Bondarev05,Bondarev06}, while conveniently representing the coupled TLS--CN system in terms of the relative distance dependent DOS functions $\xi^{\perp(\parallel)}(\mathbf{r}_{A},\omega)$. For short TLS--CN separation distances EM retardation effects play no role~\cite{Bondarev06}, and so one has $\xi^\perp\!=\xi^\parallel\!=\xi(\mathbf{r}_{A},\omega)$ for the DOS functions in Eq.~(\ref{gpm}). Figure~\ref{fig1}(b) shows $\xi(r_A\!=\!R_{cn}\!+2b,x)$, where $x=\hbar\omega/2\gamma_0$ is the dimensionless energy, $\gamma_0\!=\!2.7$~eV and $b\!=\!1.42$~\AA\space are the C-C overlap integral and interatomic distance, respectively, calculated for the three semiconducting CNs of increasing diameter. We see the sharp single-peak resonances originating from the inter-band plasmons of the respective CNs [cf. Fig.~\ref{fig1}(a) and Fig.~\ref{fig1}(b)]. These are responsible for the CN--TLS coupling in the near-field. The coupling is due to the virtual (vacuum-type) EM energy exchange between the TLS and the CN to create and annihilate plasmon excitations on the CN surface with the TLS de-excited and excited, respectively, as described by the second line in the Hamiltonian~(\ref{Htwolev}).

\section{Eigen states spectrum}

To proceed with the Raman scattering cross-section calculations, it is necessary to determine the spectrum of the eigen states of the Hamiltonian~(\ref{Htwolev}). In the linear coupling regime, quite generally, the coupled CN--TLS system can be represented as a four-level system with the eigenvectors of the Hamiltonian (\ref{Htwolev}) of the form
\begin{eqnarray}
|0\rangle\!\!&=&\!\!|l\rangle|\{0\}\rangle,\nonumber\\
|1,2\rangle\!\!&=&\!\!C_u^{(1,2)}|u\rangle|\{0\}\rangle\!+\!\!\int_{0}^{\infty}\!\!\!\!\!\!\!d\omega\!\!\int\!\!d\mathbf{R}\,C_l^{(1,2)}(\mathbf{R},\omega)
|l\rangle|\{1(\mathbf{R},\omega)\}\rangle,\nonumber\\
|3\rangle\!\!&=&\!\!|u\rangle|\{1(\mathbf{R},\omega)\}\rangle.\label{4states}
\end{eqnarray}
Here, $|\{0\}\rangle$ and $|\{1(\mathbf{R},\omega)\}\rangle$ are, respectively, the vacuum and single-quantum excited states of the CN field subsystem, and $C_{u,l}^{(1,2)}$ are unknown mixing coefficients for the non-radiative spontaneous decay transition $|u\rangle|\{0\}\rangle\!\rightarrow\!|l\rangle|\{1(\mathbf{R},\omega)\}\rangle$ to excite one plasmon of frequency $\omega$ at point $\mathbf{R}$ of the CN surface with simultaneous de-excitation of the TLS~\cite{Bondarev04,Bondarev06}. These mixing coefficients can be found by solving the eigenvalue problem for the Hamiltonian (\ref{Htwolev}) in the basis~(\ref{4states}). Similar mixing of the $|l\rangle|\{0\}\rangle$ and $|u\rangle|\{1(\mathbf{R},\omega)\}\rangle$ states, known to be responsible for the long-range dispersive van der Waals interaction~\cite{Bondarev05,Bondarev06}, is neglected here for simplicity.

Solving the eigenvalue problem for the Hamiltonian (\ref{Htwolev}) in the basis (\ref{4states}), one obtains the energy eigenvalues
\begin{equation}
E_0=-\frac{\hbar\tilde{\omega}_{A}}{2}\,,\hskip0.5cm E_3=\frac{\hbar\tilde{\omega}_{A}}{2}+\hbar\omega
\label{E0E3}
\end{equation}
for the eigenvectors $|0\rangle$ and $|3\rangle$, respectively, and the simultaneous equation set for the mixing coefficients as follows
\begin{equation}
\left\{
\begin{array}{rcl}
\displaystyle\left(\frac{\hbar\tilde{\omega}_{A}}{2}-E\!\right)C_u^{(1,2)}+
\int_{0}^{\infty}\!\!\!\!\!d\omega\!\int\!d\mathbf{R}\;\mbox{g}^{(+)}(\mathbf{r}_{A},\mathbf{R},\omega)\;C_l^{(1,2)}(\mathbf{R},\omega)\!\!\!\!\!&=&\!\!\!\!\!0\,,\\
\displaystyle\left[\mbox{g}^{(+)}(\mathbf{r}_{A},\mathbf{R},\omega)\right]^{\!\ast}\!C_u^{(1,2)}+
\left(\!-\frac{\hbar\tilde{\omega}_{A}}{2}+\hbar\omega-E\!\right)C_l^{(1,2)}(\mathbf{R},\omega)\!\!\!\!\!&=&\!\!\!\!\!0\,.
\label{eqnset}
\end{array}
\right.
\end{equation}
Here, the second equation gives
\begin{equation}
C_l^{(1,2)}(\mathbf{R},\omega)=\frac{\displaystyle\left[\mbox{g}^{(+)}(\mathbf{r}_{A},\mathbf{R},\omega)\right]^{\!\ast}}{\hbar\tilde{\omega}_{A}/2-\hbar\omega+E}\,C_u^{(1,2)}
\label{Cl}
\end{equation}
which, being inserted into the first one, results in the integral equation
\begin{eqnarray}
E\!\!\!\!\!&=&\!\!\!\!\!\frac{\hbar\tilde{\omega}_{A}}{2}+\int_{0}^{\infty}\!\!\!\!\!d\omega\!\int\!d\mathbf{R}\;\frac{\displaystyle|\mbox{g}^{(+)}(\mathbf{r}_{A},\mathbf{R},\omega)|^2}
{\hbar\tilde{\omega}_{A}/2-\hbar\omega+E}\nonumber\\
&=&\!\!\!\!\!\frac{\hbar\tilde{\omega}_{A}}{2}+\frac{\hbar^2}{2\pi}\int_{0}^{\infty}\!\!\!\!\!d\omega\;\frac{\Gamma_{0}(\omega)
(1+\omega_{A}^2/\omega^{2})\xi(\mathbf{r}_{A},\omega)}{\hbar\tilde{\omega}_{A}/2-\hbar\omega+E}
\label{eqnE12}
\end{eqnarray}
to give the energy eigenvalues $E_{1,2}$ for the eigenvectors $|1,2\rangle$. Here, the second line was obtained by using Eq.~(\ref{gpm}) with $\xi^\perp\!=\xi^\parallel\!=\xi(\mathbf{r}_{A},\omega)$ on assumption of negligible EM retardation effects at short TLS--CN separation distances~\cite{Bondarev06}. Taking advantage of the sharp single peak structure of the DOS function $\xi(\mathbf{r}_{A},\omega)$  [cf. Figs.~\ref{fig1}(a) and \ref{fig1}(b)] in the vicinity of the plasmon resonance frequency $\omega_p$~\cite{Endnote}, one can use the Lorentzian approximation of the half-width-at-half-maximum $\Delta\omega_0$ of the form $\xi(\mathbf{r}_{A},\omega)\approx\xi(\mathbf{r}_{A},\omega_p)\Delta\omega_{0}^2/[(\omega-\omega_{p})^2+\Delta\omega_{0}^2]$ to solve Eq.~(\ref{eqnE12}) analytically. One has
\[
\int_{0}^{\infty}\!\!\!\!\!d\omega\,\frac{\Gamma_{0}(\omega)(1+\omega_{A}^2/\omega^{2})\xi(\mathbf{r}_{A},\omega)}{\hbar\tilde{\omega}_{A}/2-\hbar\omega+E}
\approx\frac{\Gamma_{0}(\omega_p)(1+\omega_{A}^2/\omega_p^{2})\xi(\mathbf{r}_{A},\omega_p)\Delta\omega_0^2}{\hbar\tilde{\omega}_{A}/2-\hbar\omega_p+E}\!\!
\int_{0}^{\infty}\!\!\!\!\!\frac{d\omega}{(\omega-\omega_{p})^2+\Delta\omega_{0}^2}\,,
\]
where the integral calculates to give $[\arctan(\omega_p/\Delta\omega_0)+\pi/2]/\Delta\omega_0$, yielding $\pi/\Delta\omega_0$ with the $\arctan$ function expanded to linear terms in $\Delta\omega_0/\omega_p$ ($\ll\!1$, and the stronger this inequality is, the better such a series expansion works). Equation~(\ref{eqnE12}) now becomes a simple quadratic equation to bring one, along with Eq.~(\ref{E0E3}), to the complete energy eigenvalue set of the problem as follows
\begin{equation}
\varepsilon_0=-\frac{\tilde{x}_A}{2}\,,\;\;\;\varepsilon_{1,2}=\frac{1}{2}\left(x_p\mp\sqrt{\delta^2+X^2}-i\Delta x_p\right),\;\;\;
\varepsilon_3=\frac{\tilde{x}_A}{2}+x_p-i\Delta x_p\,.
\label{4en}
\end{equation}
Here, $\varepsilon_i=E_i/2\gamma_0$ with $i\!=\!0,1,2,3$ and $(\tilde{x}_A,x_p,\Delta x_p)=\hbar(\tilde{\omega}_A,\omega_p,\Delta\omega_p)/2\gamma_0$ are dimensionless energies, $\delta\!=\!\tilde{x}_A-x_p$, $X\!=\!(\hbar/2\gamma_0)[\,2\Delta\omega_0\Gamma_0(\omega_p)(1+\omega_A^2/\omega_p^2)\,\xi(\mathbf{r}_{A},\omega_p)]^{1/2}$, and~$\Delta x_p$ is added to phenomenologically account for the finite half-width of the plasmon resonance (finite plasmon lifetime), as seen in Fig.~\ref{fig1}(a), which is assumed to be much broader than the excited atomic level natural half-width dropped here on this account for simplicity.

\begin{figure}
\epsfxsize=8.8cm\centering{\epsfbox{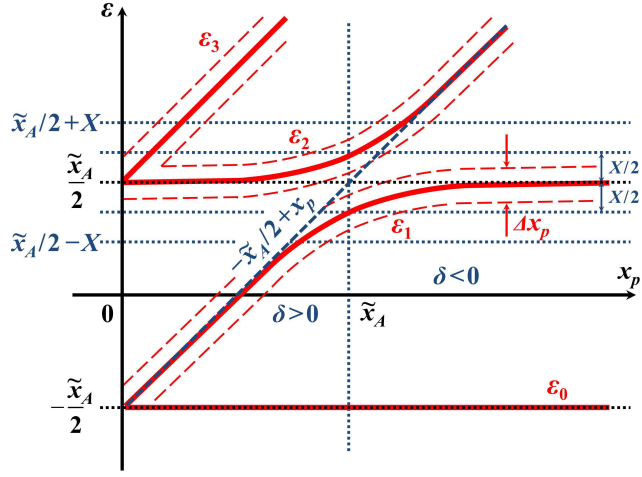}}\caption{Schematic of the energy level structure as given by Eq.~(\ref{4en}) for the coupled four-level CN--TLS system. Thick red lines show the eigen energy levels as functions of $x_p$. Thin red dashed lines indicate their broadening due to finite $\Delta x_p$. Horizontal dotted lines are to show Rabi-splitting and the in-resonance strong-coupling solutions given by Eq.~(\ref{4en}) with $\delta=0$.}\label{fig1-1}
\end{figure}

The mixing coefficients in the eigenvectors $|1\rangle$ and $|2\rangle$ in Eq.~(\ref{4states}) can be found straightforwardly by using the normalization condition
\[
|C_u^{(1,2)}|^2+\int_{0}^{\infty}\!\!\!\!\!d\omega\!\!\int\!d\mathbf{R}\,|C_l^{(1,2)}(\mathbf{R},\omega)|^2=1\,,
\]
and substituting Eq.~(\ref{Cl}) in it with $E$ replaced by $E_1$ and $E_2$, as given by Eq.~(\ref{4en}), for $C_l^{(1)}$ and $C_l^{(2)}\!$, respectively, followed by the integral evaluation within the same Lorentzian approximation for the DOS function $\xi(\mathbf{r}_{A},\omega)$ that was used to evaluate the integral in Eq.~(\ref{eqnE12}). This results in
\begin{eqnarray}
C_u^{(1,2)}\!\!\!\!\!\!\!&=&\!\!\!\!\!\!\!\left[\frac{1}{2}\left(\!1+\frac{1\mp\sqrt{1+X^2/\delta^2}}{1+X^2/\delta^2\mp\sqrt{1+X^2/\delta^2}}\!\right)\right]^{1/2}\!\!\!,\nonumber\\
\int_{0}^{\infty}\!\!\!\!\!d\omega\!\!\int\!d\mathbf{R}\,|C_l^{(1,2)}(\mathbf{R},\omega)|^2\!\!\!\!\!\!\!&=&\!\!\!\!\!\!\!1-|C_u^{(1,2)}|^2=
\frac{(X^2/2\delta^2)|C_u^{(1,2)}|^2}{1+X^2/2\delta^2\mp\sqrt{1+X^2/\delta^2}}\,.
\label{C12sqd}
\end{eqnarray}

Equations (\ref{4states}), (\ref{4en}) and (\ref{C12sqd}) represent the complete solution to the eigenvalue problem for the Hamiltonian (\ref{Htwolev}) of the coupled CN--TLS system in absence of external EM radiation. They are valid both in resonance, where $\delta\!\sim\!0$ and so $X^2/\delta^2\!\gg\!1$, and out of resonance where $X^2/\delta^2\!\ll\!1$, and give different easily derivable asymptotical expressions in these two regimes of relevance to strong and weak CN--TLS coupling, respectively. The eigen energy level structure given by Eq.~(\ref{4en}) is sketched in Fig.~\ref{fig1-1}. We see the anti-crossing behavior of the eigen energy levels $\varepsilon_1$ and $\varepsilon_2$, a characteristic of the strong coupling regime, that is controlled by the parameters $X$ and $\Delta x_p$. When in resonance, the actual coupling regime, strong or weak, depends on the relation between $X$ and $\Delta x _p$. The CN--TLS system will only be coupled strongly if the parameter $X$, which plays the role of the vacuum-field Rabi splitting here~\cite{Bondarev04,Bondarev06,Bondarev06PRB}, is much greater than the plasmon resonance broadening $\Delta x_p$ as shown in Fig.~\ref{fig1-1}. When $X\!\gg\!\Delta x_p$, the Rabi splitting of the levels $\varepsilon_1$ and $\varepsilon_2$ is not hidden by the plasmon resonance broadening, and so the strong CN--TLS coupling regime is realized. Otherwise, if $X\!\ll\!\Delta x_p$, the levels $\varepsilon_1$ and $\varepsilon_2$ are smeared, showing no clear anti-crossing behavior, and so no strong coupling can be realized.

\section{Raman scattering cross-section}

\begin{figure}
\epsfxsize=11.5cm\centering{\epsfbox{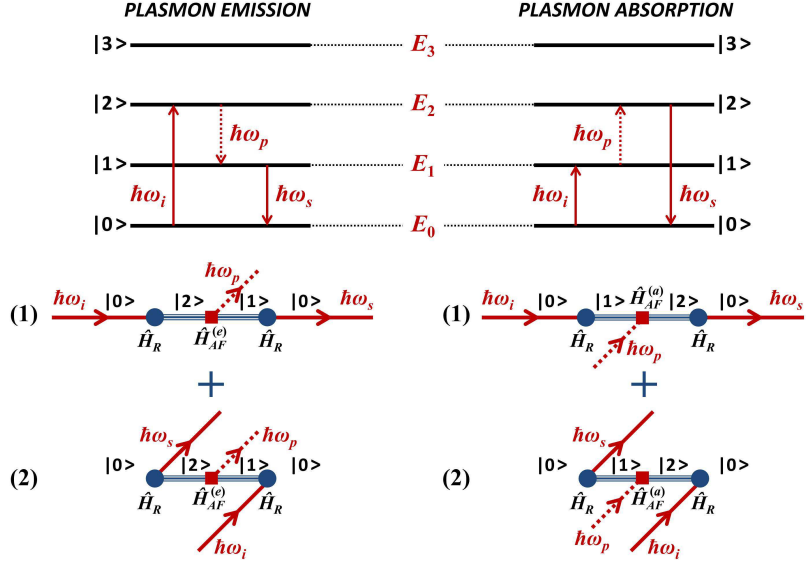}}\vskip-0.75cm\caption{Schematic of the Raman scattering process (top) in terms of the inter-level transitions (levels sketched in Fig.~\ref{fig1-1}) of the coupled CN--TLS system given by Eqs.~(\ref{4states}), (\ref{4en}) and (\ref{C12sqd}), and the Feynman diagrams (bottom) for the scattering cross-section calculations.}\label{fig2}
\end{figure}

Under the assumption that the coupled CN--TLS system with the eigen states (\ref{4states}), (\ref{4en}) and (\ref{C12sqd}) is initially in the ground state, the inelastic scattering of external EM radiation by this system only involves transitions between levels $|0\rangle$, $|1\rangle$ and $|2\rangle$, as shown in Fig.~\ref{fig2}, top, due to the dipole moment selection rule restrictions. The entire scattering process includes three sequential steps. They are:

(a)~excitation of the system by an incident photon of the frequency $\omega_i$ with the unit polarization vector $\mathbf{e}_i$, described by the interaction matrix element
\begin{equation}
\langle n|\hat{H}_R(\omega_i)|0\rangle=-\frac{i}{c}\sqrt{2\pi\hbar\omega_i}\;d_z\cos\vartheta_i\;C_u^{(n)\ast},\;\;\;
\cos\vartheta_i\!=\mathbf{e}_i\!\cdot\mathbf{e}_z\,,\;\;\;n\!=\!1,2
\label{HR}
\end{equation}
(normalized at one photon per unit volume~\cite{Berest});

(b)~plasmon emission (or absorption) on the CN surface, described by the matrix element
\begin{equation}
\langle 1|\hat{H}^{(e)}_{AF}|2\rangle=\int_{0}^{\infty}\!\!\!\!\!d\omega\!\int\!d\mathbf{R}\left[C_l^{(1)}(\mathbf{R},\omega)\,
\mbox{g}^{(+)}(\mathbf{r}_{A},\mathbf{R},\omega)\right]^{\!\ast}C_u^{(2)}
\label{HAF}
\end{equation}
[or $\langle 2|\hat{H}^{(a)}_{AF}|1\rangle=\langle 1|\hat{H}^{(e)}_{AF}|2\rangle^\dagger$ for absorption], with $\hat{H}^{(e)}_{AF}$ and $\hat{H}^{(a)}_{AF}$ being the emission term ($\sim\!\hat{f}^\dagger\,$) and the absorption term ($\sim\!\hat{f}\,$), respectively, of the interaction Hamiltonian $\hat{H}_{AF}$ in Eq.~(\ref{Htwolev});

(c)~de-excitation of~the CN--TLS system by means of the scattered (Raman) photon emission of the frequency $\omega_s$ with the unit polarization vector $\mathbf{e}_s$, described by the interaction matrix element $\langle n|\hat{H}_R(\omega_i)|0\rangle^\dagger|_{i\rightarrow s}$ in accordance with Eq.~(\ref{HR}).

There are four Feynman diagrams, shown in Fig.~\ref{fig2}, bottom, to contribute to this three-step process. They are two for plasmon emission (bottom left) and two for plasmon absorption (bottom right), to represent two indistinguishable ways for emission and absorption to occur. Two types of the emission (absorption) diagrams should be summed up and squared, followed by adding the emission and absorption contributions together~\cite{Cardona}, to result in the Fermi Golden Rule transition rate
\begin{eqnarray}
&&\hskip-1cm\left(\frac{2\pi}{\hbar}\right)\left|\frac{\langle 0|\hat{H}_R(\omega_s)|1\rangle\langle 1|\hat{H}^{(e)}_{AF}|2\rangle\langle 2|\hat{H}_R(\omega_i)|0\rangle}{[\hbar\omega_i-\hbar\omega_p-(E_1-E_0)][\hbar\omega_i-(E_2-E_0)]}\right.\nonumber\\
&&\left.+\frac{\langle 0|\hat{H}_R(\omega_i)|1\rangle\langle 1|\hat{H}^{(e)}_{AF}|2\rangle\langle 2|\hat{H}_R(\omega_s)|0\rangle}{[-\hbar\omega_s-\hbar\omega_p-(E_1-E_0)][-\hbar\omega_s-(E_2-E_0)]}\right|^2\delta(\hbar\omega_i-\hbar\omega_p-\hbar\omega_s)\nonumber\\
&&\hskip1cm+\left|\frac{\langle 0|\hat{H}_R(\omega_s)|2\rangle\langle 2|\hat{H}^{(a)}_{AF}|1\rangle\langle 1|\hat{H}_R(\omega_i)|0\rangle}{[\hbar\omega_i+\hbar\omega_p-(E_2-E_0)][\hbar\omega_i-(E_1-E_0)]}\right.\label{GRrate}\\
&&\hskip2cm\left.+\frac{\langle 0|\hat{H}_R(\omega_i)|2\rangle\langle 2|\hat{H}^{(a)}_{AF}|1\rangle\langle 1|\hat{H}_R(\omega_s)|0\rangle}{[-\hbar\omega_s+\hbar\omega_p-(E_2-E_0)][-\hbar\omega_s-(E_1-E_0)]}\right|^2\delta(\hbar\omega_i+\hbar\omega_p-\hbar\omega_s)\,.\nonumber
\end{eqnarray}
Matrix elements in here can be consistently evaluated within the Lorentzian approximation for the DOS function $\xi(\mathbf{r}_{A},\omega)$. Substituting $C_l^{(1)}$ out of Eq.~(\ref{Cl}), with $E$ replaced by $E_1$ per Eq.~(\ref{4en}), into Eq.~(\ref{HAF}) and performing exactly the same integral evaluation as was done in Eq.~(\ref{eqnE12}), one has
\[
\langle 1|\hat{H}^{(e)}_{AF}|2\rangle=\frac{2\gamma_0\;C_u^{(1)\ast}C_u^{(2)}(X^2\!/4)}{\tilde{x}_{A}/2-x_p+\varepsilon_1}=\langle 2|\hat{H}^{(a)}_{AF}|1\rangle^\dagger,
\]
yielding in view of Eq.~(\ref{HR})
\begin{eqnarray}
&&\hskip-1.3cm\langle 0|\hat{H}_R|1\rangle\langle 1|\hat{H}^{(e)}_{AF}|2\rangle\langle 2|\hat{H}_R|0\rangle=
\left(\langle 0|\hat{H}_R|2\rangle\langle 2|\hat{H}^{(a)}_{AF}|1\rangle\langle 1|\hat{H}_R|0\rangle\right)^\dagger\nonumber\\
&=&\!\!\!\!\!\frac{2\pi\hbar\sqrt{\omega_i\omega_s}}{c^2}\,d_z^2\cos\vartheta_i\cos\vartheta_s\,\frac{2\gamma_0\;|C_u^{(1)}C_u^{(2)}|^2\,(X^2\!/4)}{\tilde{x}_{A}/2-x_p+\varepsilon_1}
\label{triple}
\end{eqnarray}
with
\begin{equation}
|C_u^{(1)}C_u^{(2)}|^2=\frac{X^2\!/4}{\delta^2+X^2}\,,
\label{Cu2Cu1sqd}
\end{equation}
according to Eq.~(\ref{C12sqd}).

To obtain the differential scattering cross-section, one has to multiply Eq.~(\ref{GRrate}) by the density of final states $(\hbar\omega_s)^2d(\hbar\omega_s)d\Omega_s/(2\pi\hbar)^3$ for photons scattered within the solid angle $d\Omega_s$, and then integrate it over $\hbar\omega_s$, the scattered radiation energy~\cite{Berest}. With Eqs.~(\ref{4en}), (\ref{triple}) and (\ref{Cu2Cu1sqd}), this eventually results in the differential Raman scattering cross-section as follows
\begin{equation}
\frac{d\sigma}{d\Omega_s}\!=\!\frac{(2\gamma_0)^2|d_z|^4}{\hbar^4c^4}\cos^2\!\vartheta_{i}\,\cos^2\!\vartheta_{s}\,P(x_i,x_s)\,,
\label{crsec}
\end{equation}
with the dimensionless (angle-free) scattering probability function
\begin{eqnarray}
P(x_i,x_s)\hskip-0.2cm&=&\hskip-0.2cm x_ix_{\!s}^3\,A(\delta,X,\Delta x_p)\left\{\!\frac{1}{[(x_i-x_p-\delta_{+}/2)^2+\Delta x_p^2][(x_{\!s}-x_p-\delta_{-}/2)^2+\Delta x_p^2]}\right.\nonumber\\
&+&\hskip-0.3cm\left.\frac{1}{[(x_i-x_p-\delta_{-}/2)^2+\Delta x_p^2][(x_{\!s}-x_p-\delta_{+}/2)^2+\Delta x_p^2]}\!\right\},
\label{Pis}
\end{eqnarray}
where $x_{i,s}\!=\!\hbar\omega_{i,s}/2\gamma_0$,
\begin{equation}
A(\delta,X,\Delta x_p)=\frac{X^8}{2^6(\delta^2+X^2)^2(\delta_{\!-}^2+\Delta x_p^2)}\,,
\label{AXdxp}
\end{equation}
$\delta_{\pm}=\delta\pm\sqrt{\delta^2+X^2}$, and only the resonant terms are left to represent the contributions from plasmon emission and absorption, respectively, while (insignificant, quasi-constant background) non-resonant terms of the transition rate (\ref{GRrate}) are dropped for brevity.

\section{Discussion}

Each term in Eq.~(\ref{Pis}) has the product of \emph{two} resonance energy denominators that include $x_i$ and $x_s$, incident (incoming) and scattered (outgoing) photon energies. This is what makes the Raman scattering cross-section (\ref{crsec}) resonant. In addition to that, there is an important pre-factor there, $A(\delta,X,\Delta x_p)$, given by Eq.~(\ref{AXdxp}). This comes from Eq.~(\ref{triple}), which can be viewed as $[d_zE_z(\omega_i)][d_zE_z(\omega_s)][d_zE_z^{(loc)}(\textbf{r}_A)]^2$ with $d_zE_z^{(loc)}(\textbf{r}_A)\!\sim\!X\!\propto[\Gamma_0(\omega_p)\xi(\mathbf{r}_{A},\omega_p)]^{1/2}\!$, thus bringing the local-field enhancement factor $[d_zE_z^{(loc)}(\textbf{r}_A)]^4\propto\xi^2(\textbf{r}_A,\omega_p)$ into the Raman cross-section due to plasmon generated quasi-static electric fields (see~\cite{Bondarev12,Bondarev12pss} for more details) at the TLS location $\textbf{r}_A$ when in the CN near-surface zone $r_A\!\sim\!R_{CN}$ [Fig.~\ref{fig1}(b)]. This factor is significant when $\delta\!\sim\!0$ and $X\!\gg\!\Delta x_p$ simultaneously, a regime whereby the CN--TLS system couples strongly by means of the virtual EM energy exchange between the TLS and the nanotube, corresponding to non-exponential spontaneous decay dynamics with Rabi oscillations of the TLS excited state~\cite{Bondarev04,Bondarev06}, and Rabi splitting of the TLS optical absorption line profile~\cite{Bondarev06PRB}.

\begin{figure}
\epsfxsize=12.75cm\centering{\epsfbox{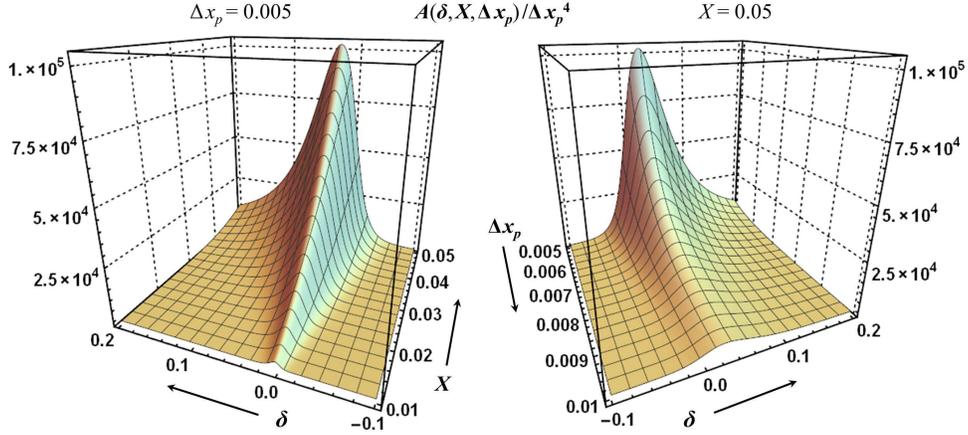}}\vskip-1.75cm\caption{Ratio $A(\delta,X,\Delta x_p)/\Delta x_p^4$ as given by Eq.~(\ref{AXdxp}) for $\Delta x_p\!=\!0.005$ (left) and for $X\!=\!0.05$ (right) to show the influence of the detuning $\delta\,(=\!\tilde{x}_A\!-\!x_p)$ on the maximum intensity of the plasmon enhanced Raman scattering effect.}\label{fig4}
\end{figure}

Maximum Raman intensity is controlled by the ratio $A(\delta,X,\Delta x_p)/\Delta x_p^4$ as can be seen from Eq.~(\ref{Pis}). This is a slightly asymmetric function of $\delta$ peaked at $\delta\!\approx\!X/\!\sqrt{3}$ when $X\!\gg\!\Delta x_p$, not at $\delta\!=\!0$ as one would expect, which can be easily shown by testing it for maximum analytically. Bondarev and Vlahovic have shown earlier in~\cite{Bondarev06PRB} by analyzing the absorption line shape profile that $X$, the Rabi splitting, may typically be as large as $\sim\!0.01-0.1$, whereas $\Delta x_p\sim\!0.005-0.01$ as can be seen from Fig.~\ref{fig1}(a). Figure~\ref{fig4}, left and right, shows the ratio $A/\Delta x_p^4$ as a function of $\delta$ and $X$ at $\Delta x_p=0.005$, and as a function of $\delta$ and $\Delta x_p$ at $X=0.05$, respectively. We see that as long as $X\!\gg\!\Delta x_p$, whereby the CN--TLS system is in the strong coupling regime, the maximum intensity decreases by a factor $\sim\!1.5$ within the detuning window $\sim\!0.08$, thus providing the spectral band as large as $0.08\!\times\!2\gamma_0\sim0.43$~eV for a significant plasmon enhanced Raman scattering effect to occur. This spectral band can be shifted both to the red and to the blue by shifting the inter-band plasmon resonance energy, as can be seen in Fig.~\ref{fig1}, which can merely be achieved by using CNs with different diameters and/or chiralities, thereby offering the flexibility and precise tunability in designing CN based SERS substrates with parameters required on-demand.

\begin{figure}
\epsfxsize=12.75cm\centering{\epsfbox{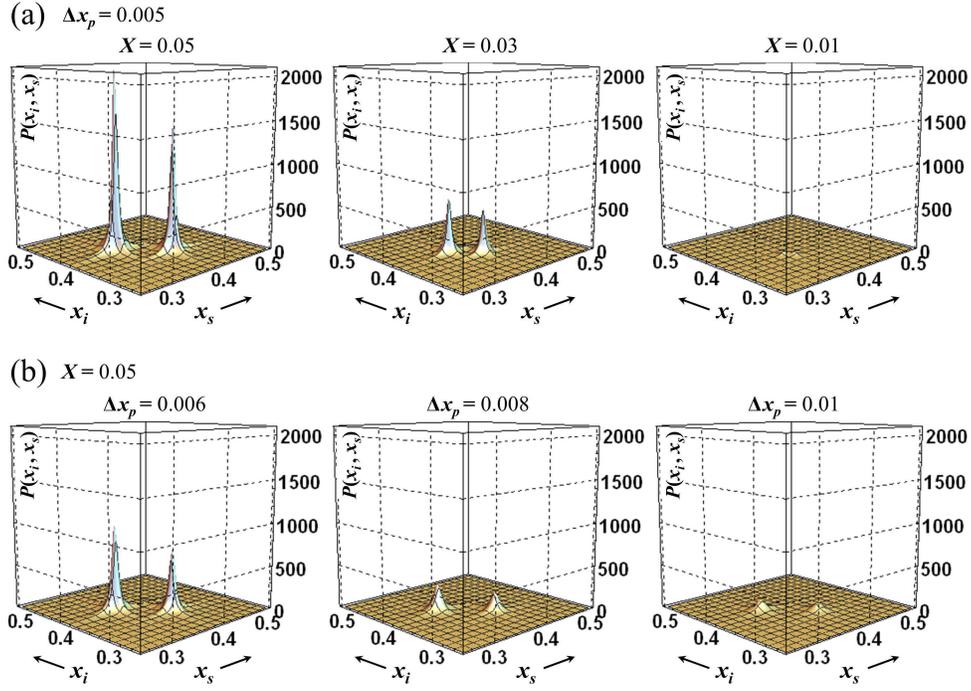}}\caption{Raman scattering probability as a function of the incident $x_i$ and scattered $x_s$ photon energies as given by Eq.~(\ref{Pis}) with $\delta\!=\!X/\!\sqrt{3}$ (maximum of $A/\Delta x_p^4$ in Fig.~\ref{fig4}, see text) for $x_p\!=\!0.35$ [$P_{11}^{(6,4)}$ plasmon in Fig.~\ref{fig1}(a)], with $X$ and $\Delta x_p$ being varied independently [rows (a) and~(b)]. TLS--plasmon coupling strength is represented by the ratio $X/\Delta x_p$ being greater or less than unity for strong and weak coupling, respectively. Raman scattering is seen to be manifestly indicative of the strong TLS--plasmon coupling, dramatically increasing as $X/\Delta x_p$ goes much greater than unity and disappearing when it is comparable with unity.}\label{fig3}
\end{figure}

Returning back to Eq.~(\ref{Pis}) we see that for each $x_i=x_p+\delta_\pm/2$ only one term contributes, resulting in either Stokes scattering with $x_s\!=\!x_p+\delta_-/2<x_i\!=\!x_p+\delta_+/2$ and a plasmon created in the CN, or in anti-Stokes scattering with $x_s\!=\!x_p+\delta_+/2>x_i\!=\!x_p+\delta_-/2$ and a plasmon absorbed from the CN. The absolute value of the Raman shift is $(\delta_+\!-\delta_-)/2\!=\!\sqrt{\delta^2\!+\!X^2}$, yielding a quantity $\sim\!X$ in resonance, where $\delta\!\sim\!0$ whereby $X^2\!/\delta^2\!\gg\!1$, and that $\sim\!\delta$ out of resonance with $X^2\!/\delta^2\!\ll\!1$. In the latter case, $A(\delta\!\gg\!X,\Delta x_p)\propto\!X^4\!/\delta^4$, according to Eq.~(\ref{AXdxp}), totally ruling out the probability $P(x_i,x_s)$ of the scattering process. When in resonance, on the other hand, the $P(x_i,x_s)$ maximum value goes as $A(0,X,\Delta x_p)/\Delta x_p^4\propto\!(X^4\!/\Delta x_p^4)/(X^2\!+\!\Delta x_p^2)$, being strongly suppressed under the weak CN--TLS coupling where $X^2\!/\Delta x_p^2\!\ll\!1$, and being dramatically enhanced in the case where $X^2\!/\Delta x_p^2\!\gg\!1$ so that the CN--TLS coupling is strong. The scattering enhancement factor is about square of that for resonance absorption by atomically doped CNs~\cite{Bondarev06PRB}, the way it should be for scattering as a two-step process of absorption followed by emission (viewed as "reversed absorption").

Figure~\ref{fig3} shows an example of the numerical calculations for the scattering probability $P(x_i,x_s)$ as given by Eq.~(\ref{Pis}) for $x_p\!=\!0.35$ [$P_{11}^{(6,4)}$ plasmon in Fig.~\ref{fig1}(a), corresponding to the energy $0.35\times2\gamma_0\!=\!1.89$~eV (red spectral line)] with $X$ and $\Delta x_p$ being varied independently [rows (a) and~(b), respectively], to see the role of the DOS resonance variation due to the local field enhancement/dehancement effect as the TLS--CN separation distance changes and the plasmon decoherence effect, respectively. As discussed above, Raman scattering is seen to be very sensitive to the strong CN--TLS coupling, blowing up by a factor of over $10^3$ for $X/\Delta x_p\!\sim\!10$ and totally vanishing when $X/\Delta x_p\!\sim\!1$.~Raising $X$ increase both the Raman shift and the intensity, while greater $\Delta x_p$ quench the intensity with no Raman shift change.

The QED theory of the plasmon enhanced Raman scattering developed here applies to chemically inactive atoms, ions, or even organic molecules and semiconductor quantum dots that are \emph{physisorbed} on the nanotube walls, whereby there is no local electronic orbitals hybridization between the nanotube and the atomic transition levels involved. The theory can be tested experimentally by using rear-earth ion complexes, Eu$^{3+}$ ions, in particular~\cite{Gaponenko99,Sandoghdar02,Noginova08}. These are known to be excellent probes for near-field effects in spatially confined systems, owing to the dominant narrow, easily detectible $5D_0\!\rightarrow\!7F_2$ electric dipole transition of the wavelength $\sim\!614$~nm between two deep-lying electronic levels ($f$-shell) of europium that essentially create an ideal TLS. Corresponding transition energy is $2.02$~eV, falls within the $0.43$~eV spectral band of the first inter-band plasmon resonance $P_{11}^{(6,4)}$ of the (6,4) nanotube (see Fig.~\ref{fig1}, $x_p\!=\!0.35$ corresponding to $E_p\!=\!1.89$~eV) whose calculated Raman spectrum is shown in Fig.~\ref{fig3} and was discussed above. Pre-alignment of europium doped CNs is desirable to facilitate the excitation efficiency, but is not crucially important.

\section{Conclusion}

In this article, the QED theory of the resonance Raman scattering is developed for a two-level dipole emitter --- TLS coupled to a weakly-dispersive low-energy ($\sim\!1\!-\!2$~eV) inter-band plasmon resonance of a carbon nanotube. The theory applies to atomic type species such as atoms, ions, molecules, or semiconductor quantum dots that are \emph{physisorbed} on the nanotube walls. The analytical expression derived for the Raman cross-section covers both weak and strong TLS--plasmon coupling, and shows dramatic enhancement in the strong coupling regime. Such resonance scattering is a manifestation of the general electromagnetic SERS effect, in which the enhancement is due to the plasmon-induced near-fields that affect the TLS in a close proximity to the CN surface, given that the TLS transition energy is within the spectral band of $\sim\!0.43$~eV of the corresponding inter-band plasmon resonance energy of the nanotube.

This theoretical work provides a unified description of the near-field plasmon enhancement effects that will help establish new design concepts for future generation CN based nanophotonics platforms with varied characteristics pre-defined on-demand --- due to extraordinary stability, flexibility and precise tunability of the CN electromagnetic properties by means of their diameter/chirality variation --- for single molecule/atom/ion detection, precision spontaneous emission control, and optical manipulation.

\section*{Acknowledgments}
This work is supported by DOE (DE-SC0007117). I.V.B. acknowledges hospitality of Munich Advanced Photonics Center at TU-Minuch, Germany, where this work was started, as well as fruitful discussions with its staff members, Prof. W.Domcke and Dr. M.Gelin.

\begin{thebibliography}{[99]}
\bibitem{Moscovits01}K.Kneipp, M.Moscovits, and H.Kneipp, \emph{Surface-Enhanced Raman Scattering: Physics and Applications} (Springer-Verlag, Berlin, 2006).
\bibitem{Otto05}A.Otto, {\lq\lq}The chemical (electronic) contribution to surface enhanced Raman scattering," J. Raman Spectrosc. \textbf{36}, 497--509 (2005).
\bibitem{ZhangNat13}R.Zhang, Y.Zhang, Z.C.Dong, S.Jiang, C.Zhang, L.G.Chen, L.Zhang, Y.Liao, J.Aizpurua, Y.Luo, J.L.Yang, and J.G.Hou, {\lq\lq}Chemical mapping of a single molecule by plasmon-enhanced Raman scattering," Nature \textbf{498}, 82--86 (2013).
\bibitem{Peng}M.Peng, H.Xu, and M.Shao, {\lq\lq}Ultrasensitive surface-enhanced Raman scattering based gold deposited silicon nanowires," Appl. Phys. Lett. \textbf{104}, 193103 (2014).
\bibitem{Hao}Q.Hao, S.M.Morton, B.Wang, Y.Zhao, L.Jensen, and T.J.Huang, {\lq\lq}Tuning surface-enhanced Raman scattering from graphene substrates using the electric field effect and chemical doping," Appl. Phys. Lett. \textbf{102}, 011102 (2013).
\bibitem{Lv}R.Lv, Q.Li, A.R.Botello-Mendez, T.Hayashi, B.Wang, A.Berkdemir, Q.Hao, A.L.Elias, R.Cruz-Silva, H.R. Gutierrez, Y.A.Kim, H.Muramatsu, J.Zhu, M.Endo, H.Terrones, J.-C.Charlie, M.Pan, and M.Terrones, {\lq\lq}Nitrogen-doped graphene: beyond single substitution and enhanced molecular sensing," Sci. Rep. \textbf{2}, 586 (2012).
\bibitem{Lin}D.Z.Lin, Y.P.Chen, P.J.Jhuang, J.Y.Chu, J.T.Yeh, and J.-K.Wang, {\lq\lq}Optimizing electromagnetic enhancement of flexible nano-imprinted hexagonally patterned surface-enhanced Raman scattering substrates," Opt. Express \textbf{19}, 4337--4345 (2011).
\bibitem{Chen}Y.-C.Chen, R.J.Young, J.V.Macpherson, N.R.Wilson, {\lq\lq}Silver-decorated carbon nanotube networks as SERS substrates," J. Raman Spectrosc. \textbf{42}, 1255--1256 (2011).
\bibitem{Sun}Y.Sun, K.Liu, J.Miao, Z.Wang, B.Tian, L.Zhang, {\lq\lq}Highly sensitive surface-enhanced Raman scattering substrate made from superaligned carbon nanotubes," NanoLett. \textbf{10}, 1747--1753 (2010).
\bibitem{Pimenta}D.M.Andrada, H.S.Vieira, M.M.Oliveira, A.P.Santos, L.Yin, R.Saito, M.A.Pimenta, C.Fantini, and C.A.Furtado, {\lq\lq}Dramatic increase in the Raman signal of functional groups on carbon nanotube surfaces," Carbon \textbf{56}, 235--242 (2013).
\bibitem{Saito}R.Saito, G.Dresselhaus, and M.S.Dresselhaus, \emph{Science of Fullerens and Carbon Nanotubes} (Imperial College, London, 1998).
\bibitem{Ando}T.Ando, {\lq\lq}Theory of electronic states and transport in carbon nanotubes," J. Phys. Soc. Jpn. \textbf{74}, 777--817 (2005).
\bibitem{ChemPhysSI}T.Hertel and I.V.Bondarev, eds., \emph{Photophysics of Carbon Nanotubes and Nanotube Composites} (Special Issue), Chem. Phys. \textbf{413}, 1--131 (2013).
\bibitem{Bondarev05}I.V.Bondarev and Ph.Lambin, {\lq\lq}van der Waals coupling in atomically doped carbon nanotubes," Phys. Rev. B \textbf{72}, 035451 (2005).
\bibitem{Bondarev06}I.V.Bondarev and Ph.Lambin, {\lq\lq}Near-field electrodynamics of atomically doped carbon nanotubes," in \emph{Trends in Nanotubes Research}, D.A.Martin, ed. (Nova Science, NY, 2006), Ch.~6, pp.~139--183.
\bibitem{Bondarev09}I.V.Bondarev, L.M.Woods, and K.Tatur, {\lq\lq}Strong exciton-plasmon coupling in semiconducting carbon nanotubes," Phys. Rev. B \textbf{80}, 085407 (2009).
\bibitem{BondarevNova11}I.V.Bondarev, L.M.Woods, and A.Popescu, {\lq\lq}Exciton-plasmon interactions in individual carbon nanotubes," in \emph{Plasmons: Theory and Applications}, K.N.Helsey, ed. (Nova Science, NY, 2011), Ch.~16, pp.~381-435.
\bibitem{Bondarev12}I.V.Bondarev, {\lq\lq}Single-wall carbon nanotubes as coherent plasmon generators," Phys. Rev. B \textbf{85}, 035448 (2012).
\bibitem{Bondarev12pss}I.V.Bondarev and T.Antonijevic, {\lq\lq}Surface plasmon amplification under controlled exciton-plasmon coupling in individual carbon nanotubes," Phys. Stat. Sol. C \textbf{9}, 1259--1264 (2012).
\bibitem{Bondarev14}I.V.Bondarev and A.V.Meliksetyan, {\lq\lq}Possibility for exciton Bose-Einstein condensation in carbon nanotubes," Phys. Rev. B \textbf{89}, 045414 (2014).
\bibitem{Popescu11}A.Popescu, L.M.Woods, and I.V.Bondarev, {\lq\lq}Chirality dependent carbon nanotube interactions," Phys. Rev. B \textbf{83}, 081406(R) (2011).
\bibitem{Woods13}L.M.Woods, A.Popescu, D.Drosdoff, and I.V.Bondarev, {\lq\lq}Dispersive interactions in graphitic nanostructures," Chem. Phys. \textbf{413}, 116--122 (2013).
\bibitem{Bondarev06PRB}I.V.Bondarev and B.Vlahovic, {\lq\lq}Optical absorption by atomically doped carbon nanotubes," Phys. Rev. B \textbf{74}, 073401 (2006).
\bibitem{Bondarev07}I.V.Bondarev and B.Vlahovic, {\lq\lq}Entanglement of a pair of atomic qubits near a carbon nanotube," Phys. Rev. B \textbf{75}, 033402 (2007).
\bibitem{GelinBondarev13}M.F.Gelin, I.V.Bondarev, and A.V.Meliksetyan, {\lq\lq}Monitoring bipartite entanglement in hybrid carbon nanotube systems via optical 2D photon-echo spectroscopy," Chem. Phys. \textbf{413}, 123--131 (2013).
\bibitem{GelinBondarev14}M.F.Gelin, I.V.Bondarev, and A.V.Meliksetyan, {\lq\lq}Optically promoted bipartite atomic entanglement in hybrid metallic carbon nanotube systems," J. Chem. Phys. \textbf{140}, 064301 (2014).
\bibitem{Bondarev10JCTN}I.V.Bondarev, {\lq\lq}Surface electromagnetic phenomena in pristine and atomically doped carbon nanotubes," J. Comp. Theor. Nanoscience \textbf{7}, 1673--1687 (2010).
\bibitem{Bondarev14Dekker}I.V.Bondarev, M.F.Gelin, and A.V.Meliksetyan, {\lq\lq}Tunable plasmon nanooptics with carbon nanotubes," in \emph{Dekker Encyclopedia of Nanoscience and Nanotechnology}, S.E.Lyshevski, ed. (3rd ed., CRC, NY, 2014), pp. 4989--5001.
\bibitem{Pichler98}T.Pichler, M.Knupfer, M.S.Golden, J.Fink, A.Rinzler, and R.E.Smalley, {\lq\lq}Localized and delocalized electronic states in single-wall carbon nanotubes," Phys. Rev. Lett. \textbf{80}, 4729--4732 (1998).
\bibitem{Tasaki98}S.Tasaki, K.Maekawa, and T.Yamabe, {\lq\lq}$\pi$-band contribution to the optical properties of carbon nanotubes: Effects of chirality," Phys. Rev. B \textbf{57}, 9301--9318 (1998).
\bibitem{Li01}Z.M.Li, Z.K.Tang, H.J.Liu, N.Wang, C.T.Chan, R.Saito, S.Okada, G.D.Li, J.S.Chen, N.Nagasawa, and S.Tsuda, {\lq\lq}Polarized absorption spectra of single-walled 4~\AA\space carbon nanotubes aligned in channels of an AlPO$_4$--5 single crystal," Phys. Rev. Lett. \textbf{87}, 127401 (2001).
\bibitem{Bondarev02}I.V.Bondarev, G.Ya.Slepyan, and S.A.Maksimenko, {\lq\lq}Spontaneous decay of excited atomic states near a carbon nanotube," Phys. Rev. Lett. \textbf{89}, 115504 (2002).
\bibitem{Marin03}A.G.Marinopoulos, L.Reining, A.Rubio, and N.Vast, {\lq\lq}Optical and loss spectra of carbon nanotubes: Depolarization effects and intertube interactions," Phys. Rev. Lett. \textbf{91}, 046402 (2003).
\bibitem{Bondarev04}I.V.Bondarev and Ph.Lambin, {\lq\lq}Spontaneous-decay dynamics in atomically doped carbon nanotubes," Phys. Rev. B \textbf{70}, 035407 (2004).
\bibitem{Endnote}Actual DOS resonance frequencies are slightly red shifted relative to their respective plasmon resonance frequencies [cf. Figs.~\ref{fig1}(a) and \ref{fig1}(b)]. The shifts are within plasmon resonance widths though, and so are neglected, thereby reducing the number of relevant theory parameters here.
\bibitem{Berest}V.B.Berestetskii, E.M.Lifshitz, and L.P.Pitaevskii, \emph{Qua\-n\-tum Electrodynamics} (Pergamon, Oxford, 1982).
\bibitem{Cardona}P.Y.Yu and M.Cardona, \emph{Fundamentals of Semiconductors}, 4th edn. (Springer-Verlag, Berlin, 2010).
\bibitem{Gaponenko99}S.V.Gaponenko, V.N.Bogomolov, E.P.Petrov, A.M.Kapitonov, D.A.Yarotsky, I.I.Kalosha, A.A.Eychmueller, A.L.Rogach, J.McGilp, U.Woggon, and F.Gindele, {\lq\lq}Spontaneous emission of dye molecules, semiconductor nanocrystals, and rare-earth ions in opal-based photonic crystals," J. Lightwave Technol. \textbf{17}, 2128--2138 (1999).
\bibitem{Sandoghdar02}H.Schniepp and V.Sandoghdar, {\lq\lq}Spontaneous emission of europium ions embedded in dielectric nanospheres," Phys. Rev. Lett. \textbf{89}, 257403 (2002).
\bibitem{Noginova08}N.Noginova, G.Zhu, M.Mavy, and M.A.Noginov, {\lq\lq}Magnetic dipole based systems for probing optical magnetism," J. Appl. Phys. \textbf{103}, 07E901 (2008).
\end{thebibliography}
\end{document}